\documentclass[%
preprint,
superscriptaddress,
 amsmath,amssymb,
prb,
]{revtex4-2}

\usepackage{chemformula}
\usepackage{graphicx}
\usepackage{dcolumn}
\usepackage{bm}

\begin{document}
\title{Hydrogen in tungsten trioxide by membrane photoemission and DFT modelling}

\author{Emanuel Billeter}
\affiliation{Laboratory for Advanced Analytical Technologies, Empa - Swiss Federal Laboratories for Materials Science and Technology, \"Uberlandstrasse 129, 8600 D\"ubendorf, Switzerland}
\affiliation{Department of Chemistry, University of Z\"urich, Winterthurerstrasse 190, 8057 Z\"urich, Switzerland}

\author{Andrea Sterzi}
\affiliation{Laboratory for Advanced Analytical Technologies, Empa - Swiss Federal Laboratories for Materials Science and Technology, \"Uberlandstrasse 129, 8600 D\"ubendorf, Switzerland}

\author{Olga Sambalova}
\affiliation{Laboratory for Advanced Analytical Technologies, Empa - Swiss Federal Laboratories for Materials Science and Technology, \"Uberlandstrasse 129, 8600 D\"ubendorf, Switzerland}
\affiliation{Department of Chemistry, University of Z\"urich, Winterthurerstrasse 190, 8057 Z\"urich, Switzerland}

\author{René Wick-Joliat}
\affiliation{Department of Chemistry, University of Z\"urich, Winterthurerstrasse 190, 8057 Z\"urich, Switzerland}

\author{Cesare Grazioli}
\affiliation{IOM-CNR, Laboratorio TASC, Basovizza SS-14, km 163.5, 34149 Trieste, Italy}

\author{Marcello Coreno}
\affiliation{ISM-CNR, Istituto di Struttura della Materia, LD2 Unit, 34149 Trieste, Italy}

\author{Yongqiang Cheng}
\affiliation{Neutron Scattering Division, Oak Ridge National Laboratory, Oak Ridge, TN 37831, USA}

\author{Anibal J. Ramirez-Cuesta}
\affiliation{Neutron Scattering Division, Oak Ridge National Laboratory, Oak Ridge, TN 37831, USA}

\author{Andreas Borgschulte}  \email{andreas.borgschulte@empa.ch}
\affiliation{Laboratory for Advanced Analytical Technologies, Empa - Swiss Federal Laboratories for Materials Science and Technology, \"Uberlandstrasse 129, 8600 D\"ubendorf, Switzerland}
\affiliation{Department of Chemistry, University of Z\"urich, Winterthurerstrasse 190, 8057 Z\"urich, Switzerland}

\date{\today}

\begin{abstract}
	The measurement of hydrogen induced changes on the electronic structure of transition metal oxides by X-ray photoelectron spectroscopy is a challenging endeavor, since no photoelectron can be unambiguously assigned to hydrogen. The H-induced electronic structure changes in tungsten trioxide have been known for more than 100 years, but are still controversially debated. The controversy stems from the difficulty in disentangling effects due to hydrogenation from the effects of oxygen deficiencies. Using a membrane approach to X-ray photoelectron spectroscopy, in combination with tunable synchrotron radiation we measure simultaneously core levels and valence band up to a hydrogen pressure of 1000 mbar. Upon hydrogenation, the intensities of the W$^{5+}$ core level and a state close to the Fermi level increase following the pressure-composition isotherm curve of bulk H$_x$WO$_3$. Combining experimental data and density-functional theory the description of the hydrogen induced coloration by a proton polaron model is corroborated. Although hydrogen is the origin of the electronic structure changes near the Fermi edge, the valence band edge is now dominated by tungsten orbitals instead of oxygen as is the case for the pristine oxide, having wider implication for its use as a (photo-electrochemical) catalyst.
\end{abstract}

\pacs{Valid PACS appear here}
\maketitle

\section{\label{sec:1}Introduction}
Hydrogen is an ubiquitous element in the environment. The element plays a key role in biology, chemistry and physics: It is involved in numerous chemical reactions, from photosynthesis to the combustion of its products and plays an essential role in corrosion processes. The fast diffusion of hydrogen in most materials including non-organic matter such as oxides, makes hydrogen an omnipresent impurity \cite{Volkl1981}. Due to its polyvalent chemical character, hydrogen in matter may be present as proton, neutral atom, or as anion.  As a consequence of its atomic number, hydrogen has only one electron. Particularly this property is a challenge for many analytical tools based on the interaction with electrons: core-level spectroscopies such as X-ray photoelectron spectroscopy (XPS) cannot be used as a quantitative method for hydrogen, because a photo-emitted electron cannot be assigned to a hydrogen atom without any doubt, H$^+$ (e.g. OH) has formally no electrons, H$^-$ has two, and in covalent bonds the H electron has a high probability density between the binding atoms, i.e., the location of hydrogen electrons depends on the electronic structure of the material. The electronic structure, though, is the key to understand the physical and chemical properties of matter, and photoemission spectroscopy is the standard experimental method to unravel it. Obviously, materials, in which the electronic structure is decisively influenced by hydrogen are particularly difficult to  analyze. The archetypal example of this materials class is hydrogen intercalated tungsten trioxide, H$_x$WO$_3$.  Already in the 19$^{th}$ century, Berzelius noticed a color change when hydrogen gas was passed over tungsten trioxide \cite{Wohler1824}. Similar reversible optical and electrical changes are observed upon electrochemical insertion of alkali metals into WO$_3$ \cite{Brimm1951}. The discovery of the electrochromic properties of thin WO$_3$ films led to the development of a number of applications such as smart windows, displays and variable mirrors \cite{Deb1973, Granqvist2000}. As the above discussed constraints for hydrogen do not apply here, the electronic structure of alkali-metal intercalated tungsten bronzes, e.g., Na$_x$WO$_3$, are well characterized \cite{Hochst1981,Raj2009}. With the help of electronic structure modeling \cite{Hjelm1996,Larsson2003},  there is consensus about the underlying physical phenomena of the chemical modification induced by alkali-metal intercalation \cite{Granqvist1995}. In short, the electron of the alkali-metal forms a new state in the band gap (conducting phase), but the phase initially remains insulating due to Coulomb interactions. In Na$_x$WO$_3$, a metal-insulator transition occurs at $x=0.24$; similar behavior is found for the other alkali-metal intercalated tungsten bronzes \cite{Hochst1982,Larsson2003}. Although intuitively similar, the changes taking place during hydrogen intercalation are different and controversially debated. Simplified, there are two models: hydrogen is intercalated into WO$_3$ and remains there, most likely as hydroxide; the corresponding electron affects the valence band in a similar way as the ones originating from alkali-metals \cite{Hjelm1996, Leng2017}. The second model proposes that  hydrogen can form water with oxygen from WO$_3$, and leave the now substoichiometric crystal \cite{Saenger2008}. The corresponding oxygen vacancies are filled with electrons. The loss of oxygen from the lattice induces distortions, leading to the localization of the electrons, that are described by an electron polaron model \cite{Johansson2016}. Summarizing, the origin of the controversy stems from the inability to detect and characterize hydrogen in the oxide.

In contrast to the electronic structure, the crystal structures of hydrogen and alkaline metal intercalated WO$_3$ are well studied. Most alkaline metal  bronzes have a perovskite structure, where the alkaline metal occupies the central position \cite{Brimm1951}. The structure of hydrogen intercalated WO$_3$ was determined by  X-ray diffraction (XRD) and neutron diffraction on deuterium analogs \cite{Dickens1967, Wiseman1973}. It revealed a distorted cubic structure, where the hydrogen occupies a position 1.1~\AA~from the oxygen along the diagonal through the central position. This was supported by DFT calculations on cubic WO$_3$ and HWO$_3$ finding the minimum energy position of hydrogen in the WO$_3$ lattice  at a distance of 1.03 \AA~ from the oxygen atoms \cite{Hjelm1996}. Electronic structures have been calculated for hydrogen and alkali metal intercalation explaining the observed changes at high intercalant concentration \cite{Hjelm1996, DeWijs1999}.

Experimentally, the intercalation process can be performed in two different ways. Electrochemical hydrogen insertion is relatively facile, but the electrochemical surface changes due to the aqueous environment hinder photoemission measurements. Gaseous hydrogen intercalation into WO$_3$ does not change the surface but is feasible only at UHV-incompatible pressures and in the presence of dissociatively active sites \cite{Berzins1983}, limiting photoemission experiments to the study of \textit{post-mortem} samples.

\begin{figure}
	\includegraphics[width=0.5\textwidth]{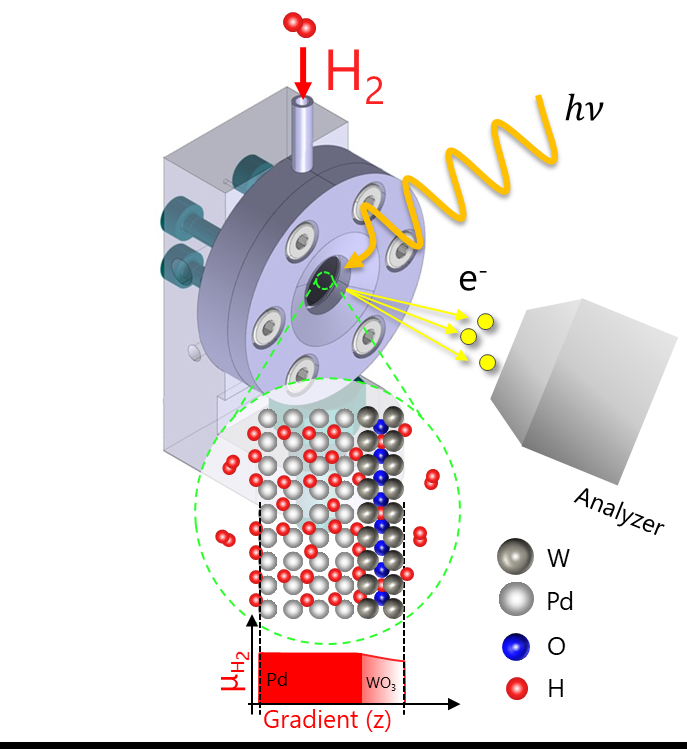}
	\caption{\label{Fig:1_sketch} Sketch of the membrane XPS sample holder and the experimental setup. The enlarged area shows the hydrogen permeation and the hydrogen chemical potential through the membrane. At steady state, the chemical potential at the feed side of the membrane is nearly equal to that at the vacuum side allowing the measurement of hydrogen concentrations in the material corresponding up to atmospheric pressures by surface science techniques \cite{Delmelle2015,Sambalova2018}.}
\end{figure}

In this study, we present \textit{in-situ} photoemission data obtained by the membrane photoemission approach \cite{Delmelle2015}: we employ a sample holder that enables varying the hydrogen pressure up to one bar while keeping the tungsten oxide thin film under UHV conditions necessary for photoelectron spectroscopy experiments \cite{Delmelle2015} (see Fig.~\ref{Fig:1_sketch}). This allows to measure the pressure-composition isotherm of WO$_3$ by photoelectron spectroscopy. Synchrotron light enables probing of oxygen and tungsten core levels and the valence band states as a function of hydrogen content. The \textit{operando} approach facilitates interpretation, as effects from unavoidable substoichiometries of the sample and other experimental uncertainties can be separated from the effects expected from hydrogen intercalation. Furthermore we compare the photoemission results with DFT calculations, supporting the interpretation of experimental results in light of a polaron model.

\section{\label{secExp} Experimental}
\subsection{\label{subsec:Sample Prep}Sample preparation}
WO$_{3}$ was deposited on palladium by electrodeposition from a H$_2$WO$_4$ solution. The H$_2$WO$_4$ solution was prepared in a round bottom flask by suspending 0.92~g tungsten powder (99.9~\%, Sigma-Aldrich) in 7.5~ml water. The suspension was heated to 60~$^\circ$C and 3.5~ml H$_2$O$_2$ (30~\%, Merck) were added. Tungsten was oxidized to tungstic acid accompanied with vigorous gas formation (H$_2$) and after 2~minutes a colorless, clear solution was obtained. A platinum wire was added to decompose the excess of H$_2$O$_2$ until no more oxygen bubbles evolved, which took 8~hours at 60~$^\circ$C. A final tungstic acid concentration of 50~mM was obtained by adding 50~ml of isopropyl alcohol and filling up with water to a total volume of 100~ml. Electrodeposition was performed in a two electrode setup. A 2x2~cm foil of Pd with 47~$\mu$m thickness (purity 99.95~\%, Goodfellow) with its backside covered with Kapton tape served as working electrode and a Pt wire as counter electrode. Since the WO$_3$ deposition on Pd competes with H$_2$ formation, relatively high current densities of approximately -2~mA/cm$^2$ were applied in chronopotentiometry mode for 10~minutes, which resulted in an amorphous blue film of approximately 30~nm thickness, as determined by XRD and EDX, respectively. The sample was then dried in air on a hot plate at 200~$^\circ$C for 10~minutes. \\

\subsection{\label{subsec:mebrane PES} Membrane sample holder}
The high pressure XPS study reported here is based on the Pd membrane approach that we previously presented \cite{Sambalova2018}. The photoelectron spectroscopy measurements were carried out at the GasPhase beamline of the Elettra Synchrotron light source in Trieste (Italy). The beamline is equipped with a dedicated differential pumping system \cite{Blyth1999}. The combination of a high resolution monochromator with the high transmission of the electron analyzer allowed to collect spectra with a minimal energy resolution of 50~meV \cite{Grazioli2017}. 

Briefly, the idea behind the embedded membrane approach is the following. The sample holder sketched in Fig. \ref{Fig:1_sketch} consists of a hydrogen permeable Pd membrane which is exposed at one side to high hydrogen pressure (up to 1 bar) and at the other side to UHV ($10^{-8}$ mbar). Hydrogen is fluxed at the high pressure side of the membrane and, following permeation, it reaches the surface exposed to vacuum in the form of atomic hydrogen. The tungsten oxide  film, which is deposited on the analyzed side of the membrane, is thereby hydrogenated. Since the hydrogen permeation in the palladium membrane is much faster than in tungsten oxide, the oxide layer blocks hydrogen permeation. This creates a constant hydrogen concentration, and therefore a constant chemical potential, in the palladium membrane \cite{Delmelle2015}. It is thereby possible to control the chemical potential of hydrogen at the UHV side of the membrane by the applied hydrogen pressure (see Fig. \ref{Fig:1_sketch}). The validity of the assumption of the equality of chemical potentials is experimentally evidenced by the permeation kinetics: the analysis gives a surface limited permeation (see appendix, Fig.~\ref{fig:doublelog}).

Photoemission measurements require this system to be in a pressure-temperature equilibrium state. In order to realize that, both the gas pressure and the membrane temperature were adjusted. The membrane was heated up to 160$~^\circ$C by means of a heating filament and the hydrogen flux was controlled through a valve and a pressure gauge. Pressure-dependent spectra were obtained as follows. Hydrogen was constantly fluxed inside the sample holder, an equilibrium condition is reached when the amount of absorbed gas compensates the desorption process which occurs at the UHV side of the membrane \cite{Sambalova2018}. When this steady state has been reached, the pressures at the two sides of the membrane are stable and the measurement can be performed. The hydrogen amount inside the experimental chamber and hence the desorption process were monitored by means of a quadrupole residual gas analyser (see appendix, Fig.~\ref{fig:doublelog}).

\subsection{\label{subsec:XPS} Photoelectron spectroscopy}
\textit{In-situ} synchrotron measurements were recorded using 700~eV photon energy for the survey spectra and 104~eV for the simultaneous detection of tungsten 4f core levels and the valence band. During hydrogenation experiments, the photoemission spectra were recorded continuously in single sweep mode, with each spectrum taking approximately 120~s. The applied hydrogen pressure was measured by an external pressure gauge. As soon as the partial hydrogen pressure in the chamber was constant, the external hydrogen pressure was increased. This lead to approximately ten spectra being recorded for each external hydrogen pressure. These spectra were averaged for the subsequent data evaluation to improve signal to noise ratios.  Data analysis was performed using the CasaXPS software, employing the GL(30) line shape, Shirley background subtraction and a 2.1 eV spin orbit splitting for both W$^{5+}$ and W$^{6+}$ doublets \cite{Bouvard2016}. The W5p$_{3/2}$ core level is not fitted in the spectra recorded at 104~eV, due to its small photoionization cross section at this photon energy \cite{Yeh1985}. The binding energy calibration was performed on the gold coated sample holder, setting the Au 4f$_{7/2}$ peak to 84.0~eV. Due to small photon energy instabilities and work function changes during hydrogenation, all spectra were subsequently aligned by shifting the W$^{6+}$ 4f$_{7/2}$ peak to 35.8~eV. These shifts are on the order of 0.2-0.3~eV.

\textit{Ex-situ} measurements were performed using a PHI \textit{Quantes} spectrometer (ULVAC-PHI) equipped with a monochromatic Al K$\alpha$ (1486.6~eV) X-ray source. Charge neutralization was accomplished by a dual beam charge neutralization system, employing low energy electron and argon ion beams. Detailed acquisition parameters are given in table \ref{tab:XPS_Parameters}.

\subsection{\label{subsec:DFT}Electronic structure calculation}
Calculation of electronic structure by Density Functional Theory (DFT) and hybrid-DFT was performed using the Vienna Ab initio Simulation Package (VASP) \cite{Kresse1993,Kresse1994,Kresse1996,Kresse1996b}. The calculation used Projector Augmented Wave (PAW) method \cite{Blochl1994,Kresse1999} to describe the effects of core electrons, and Perdew-Burke-Ernzerhof (PBE) \cite{Perdew1996}  implementation of the Generalized Gradient Approximation (GGA) for the exchange-correlation functional. Energy cutoff was 600 eV for the plane-wave basis of the valence electrons. The electronic structure was calculated on a 15x15x15 $\Gamma$-centered mesh for WO$_3$ (unit cell), and a 4x4x4 $\Gamma$-centered mesh for H$_x$WO$_3$ (2x2x2 supercell). The total energy tolerance for electronic energy minimization was 10$^{-6}$ eV, and for structure optimization it was 10$^{-5}$ eV. The maximum interatomic force after relaxation was below 0.01 eV/Å. After structural optimization, hybrid-DFT calculation of band structure was performed using the HSE06  hybrid functional \cite{Heyd2003,Krukau2006} with a mixing parameter of 25\% and a screening parameter of 0.2 \AA$^{-1}$. 

Hydrogen free WO$_3$ crystallizes in the monoclinic structure with space group P2$_1$/n with a = 7.304 \AA, b  = 7.536 \AA, c = 7.691 \AA~and $\beta$ = 90.85$^\circ$ \cite{Lassner1999}. Other polymorphs exist depending on temperature \cite{Wriedt1989}. However, intercalation of hydrogen leads to the formation of cubic structures of H$_x$WO$_3$ for $x=0.5$ \cite{Wiseman1973}. We thus simplified the calculations using the cubic ReO$_3$ structure where the metal is surrounded by six oxygen atoms in a octahedron (see Fig.~\ref{Fig:Structure}) for all compositions including hydrogen free WO$_3$. With this constraint, the calculated electronic structure is that of an artificial structure with slight deviations from reality. A simple quality parameter is the optical gap \cite{DeWijs1999,Gonzalez2010}. The calculated direct band gap of simple cubic WO$_3$ with around 2.3 eV  match the experimental values of 2.6 to 3.3 eV for the direct band gap \cite{Green1991,Saenger2008,Gonzalez2010}. This difference is due to the general underestimation of the optical gap by GGA and WDA methods \cite{Xiong2007}, with WO$_3$ being particularly notorious \cite{Wang2011}. Furthermore,  calculations of the gap of the simple cubic phase yield smaller gaps than the ones calculated for the monoclinic phase of bulk tungsten trioxide \cite{Cora1996, Hjelm1996,DeWijs1999}. Values vary from 0.69 eV (RPBE) and 2.25 eV (PBE8) for the simple cubic system to 1.3 eV (RPBE) and 3.68 eV (PBE8) for the monoclinic system with functional indicated in brackets \cite{Wang2011}. However, the simplicity of a cubic system allows the modelling of the partially hydrogenated bronzes H$_x$WO$_3$ with $x=0, 0.25,  0.75,$ and $x= 1.0$. For $x= 0.25$, two hydrogens with antiparallel spin configuration are placed in a simple cubic eightfold unit cell of WO$_3$  (see Figs.~\ref{Fig:Structure},~\ref{fig:HconstellationDFT}). Nine different hydrogen constellations were calculated, with the two hydrogen atoms placed in the unit cells along the 100, 110 and 111 direction respectively. The difference between these constellations turned out to be negligible (Appendix, Fig.~\ref{fig:HconstellationDFT}). The hydrogen and oxygen positions were released in the sub-unit to find the H-O position with minimum total energy. We found an optimum for $d_{O-H} = 1.03$~\AA , in good agreement with literature \cite{Hjelm1996}. In addition, the oxygen tetrahedrons were distorted (Fig.~\ref{Fig:Structure}). Similar calculations were performed for H$_{0.75}$WO$_3$ using 6 hydrogens per eight fold unit cell. For better comparison, the calculated  total density of electron states are broadened by 0.5~eV to match the experimental resolution of photoemission spectroscopy (see Fig. \ref{Fig:DFT_vs_Exp}).

\section{\label{sec:results}Results}
For characterization, XPS measurements employing Al-K$\alpha$ radiation were performed on the tungsten oxide coated palladium membrane pre and post hydrogenation (see Appendix Fig.~\ref{fig:Survey_spectra}). The results are in perfect agreement with literature; i.e., we find mainly W$^{6+}$ and  some W$^{5+}$ as extracted by peak fitting of the W4f states \cite{DeAngelis1977}, which is typical for amorphous WO$_3$ thin films \cite{Barreca2001}. There is no significant difference between the data taken \textit{ex-situ} before and after hydrogen exposure. The survey spectra show less carbon contamination after hydrogenation and the presence of Pd3d peaks, that are attributed to inadvertent removal of the thin WO$_3$ layer during mounting or dismounting from the membrane sample holder.

\begin{figure}
    \centering
    \includegraphics{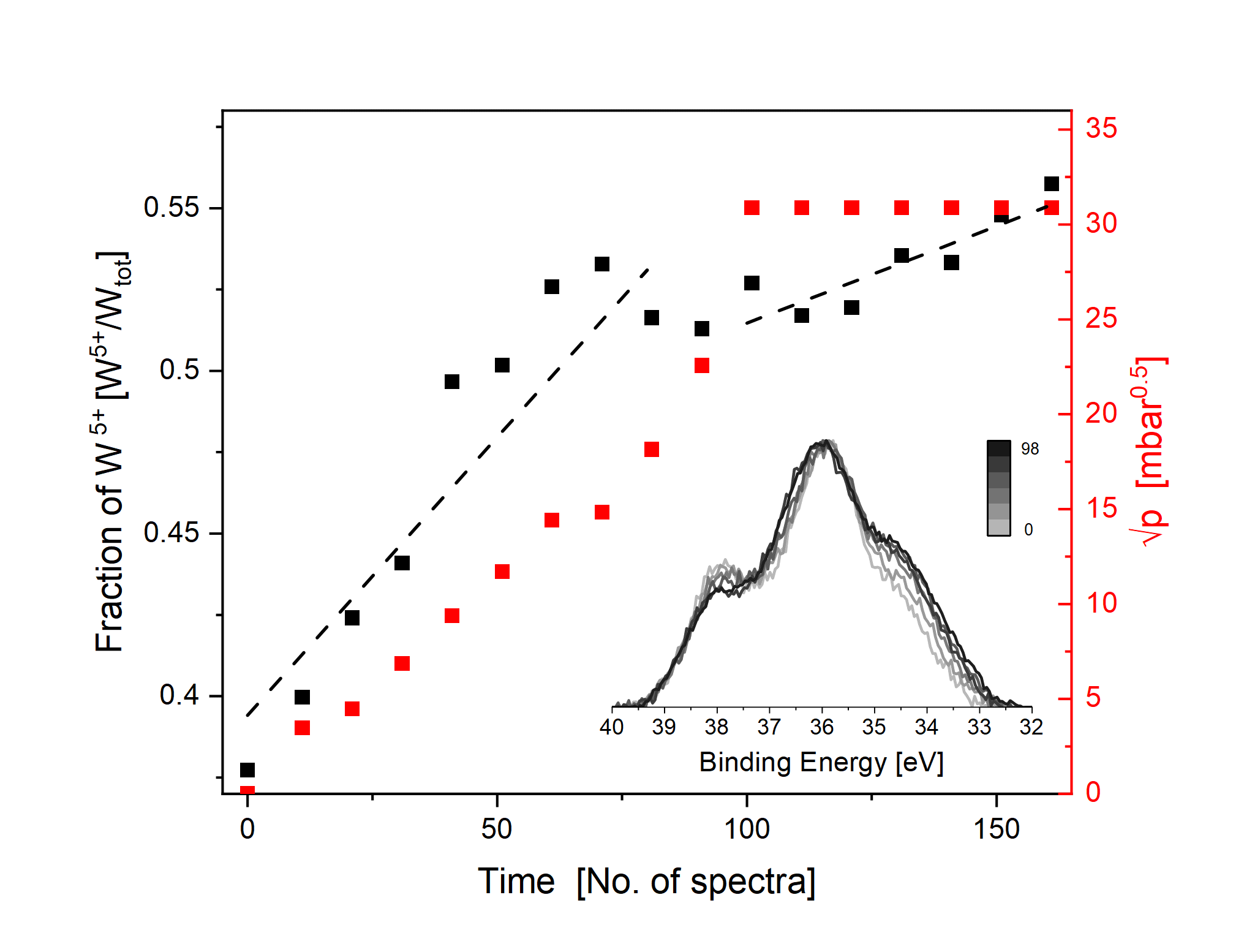}
    \caption{The fraction as derived from peak fitting shown in the appendix (Fig.~\ref{fig:Survey_spectra} ) of W$^{5+}$ in the sample is shown in black, with the linear interpolation of the slope shown by the dashed gray line during the increasing hydrogen pressure and the constant hydrogen pressure phases, shown in red. The gray scale spectra show the changes in the W4f spectra during the increasing hydrogen pressure phase.}
    \label{fig:W4f_reduction}
\end{figure}

\begin{figure}
	\includegraphics[width=0.5\textwidth]{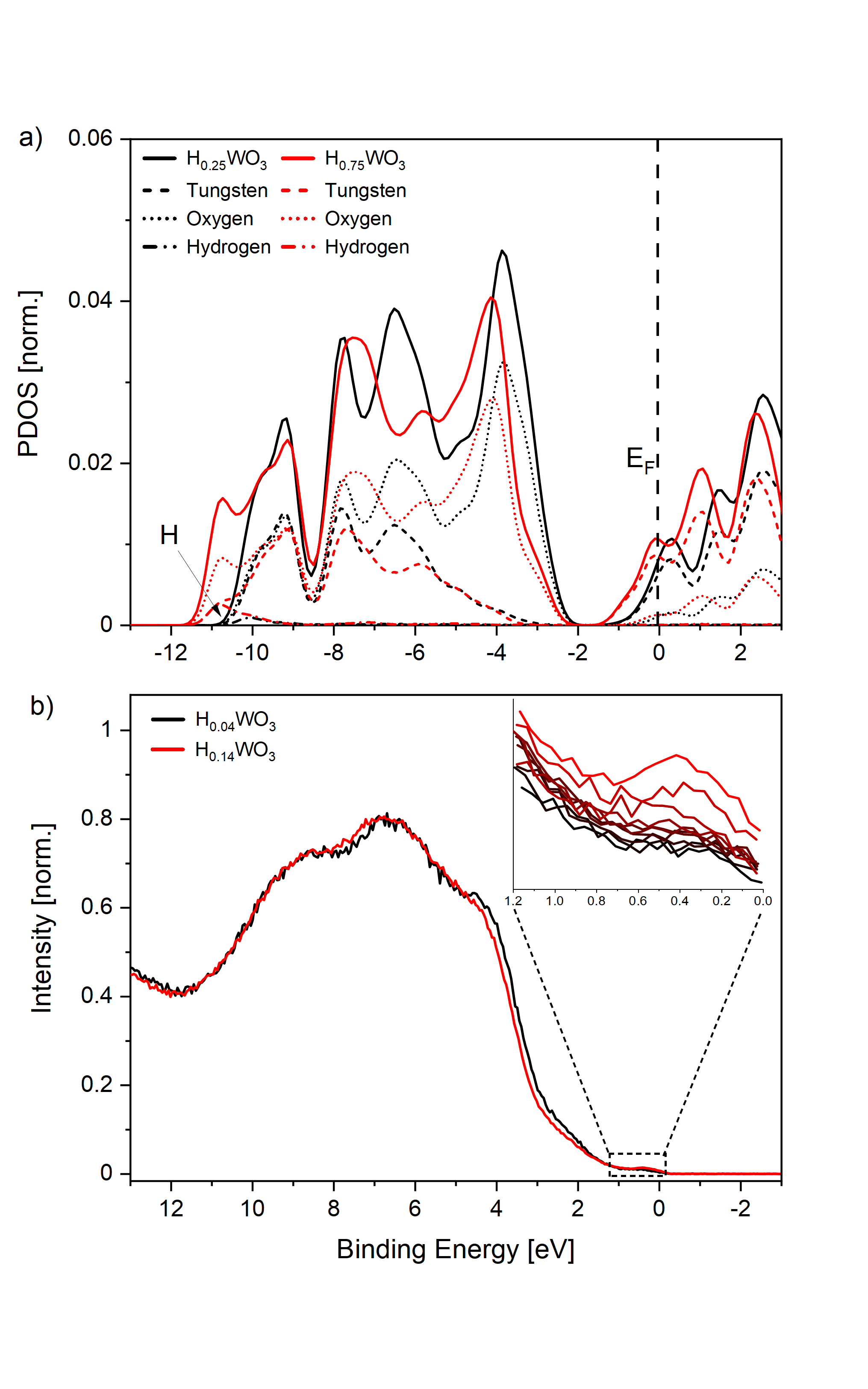}
	\caption{\label{Fig:DFT_vs_Exp}
	a) Calculated partial and total DOS of H$_{0.25}$WO$_3$ (black) and H$_{0.75}$WO$_3$ (red). b)  Measured photoemission spectra of minimum and maximum hydrogen content in black and red. The hydrogen contents as labelled are derived from the peak intensities of the peak evolving with hydrogen pressure as shown in the inset (see Fig.~\ref{fig:4_VB_H2_pressure} and text for more information).}
\end{figure}

The same procedure was applied \textit{in-situ} directly before and after hydrogen exposure (Appendix, Fig.~\ref{fig:700eV_spectra}). After hydrogen exposure the ratio W$^{5+}/W^{6+}$ increases. The intensity of the higher binding energy shoulder of the O1s peak, indicative of OH and/or carbonate species \cite{Barreca2001}, increases in parallel (Appendix, Fig.~\ref{fig:700eV_spectra}). These changes were followed  \textit{in-situ} at various hydrogen pressures applied to the membrane. We observed continuous variations of core levels (Fig.~\ref{fig:W4f_reduction}) as well as the valence band (Fig.~\ref{Fig:DFT_vs_Exp}b) indicative of hydrogen intercalation into WO$_3$. However, in addition to the effects assigned to hydrogen exposure, we found a higher W$^{5+}$/W$^{6+}$ ratio derived from \textit{in-situ} measurements at 700~eV than the one derived \textit{ex-situ} at 1486.6~eV photon energy, both measured on the sample before hydrogen exposure. This effect is attributed to the prolonged UHV exposure necessary to align the experimental setup. During alignment and  optimization of the acquisition parameters, the membrane was exposed to UV radiation of 104~eV photon energy. It is well known that UV radiation induces the reduction of WO$_3$, leading to larger fractions of W$^{5+}$ \cite{Bussolotti2003}. This effect is a major detriment of photoelectron spectroscopy applied to such systems in general. 

The  \textit{in-situ} hydrogenation opens the possibility to disentangle the effects from radiation damage and hydrogen exposure  by following the evolution of the spectra with and without hydrogen. In Fig.~\ref{fig:W4f_reduction}, the WO$_3$ thin film was exposed to increasing hydrogen back pressure up to 1~bar and then kept under these conditions for a long period. The increase of the W$^{5+}$ fraction shows two different slopes. With increasing the hydrogen back pressure, the amount of W$^{5+}$ grows fast. When the hydrogen back pressure is high but constant, the growth of W$^{5+}$ slows down markedly. The amount of W$^{5+}$ purely from hydrogenation may then be the difference between the two dashed fitting curves, neglecting a mutual enhancement of the two effects. The hydrogen content in H$_x$WO$_3$ has been shown to be proportional to the square root of the applied hydrogen pressure \cite{Berzins1983,Fripiat1992}. With this relation, the amount of W$^{5+}$ is proportional to $\sqrt{p_{H_2}}$ and thus to the amount of hydrogen $x$ in H$_x$WO$_{3-\delta}$. An uncertainty is the number of oxygen vacancies $\delta$. The measured differences of the O to W elemental ratio between pre- and post hydrogenation  is less than 3\% (Appendix, (Fig.~\ref{fig:700eV_spectra})). This lies within the accuracy of the measurements. Within this limitation, there is no oxygen loss during hydrogenation.

\begin{figure}
	\includegraphics[width=0.5\textwidth]{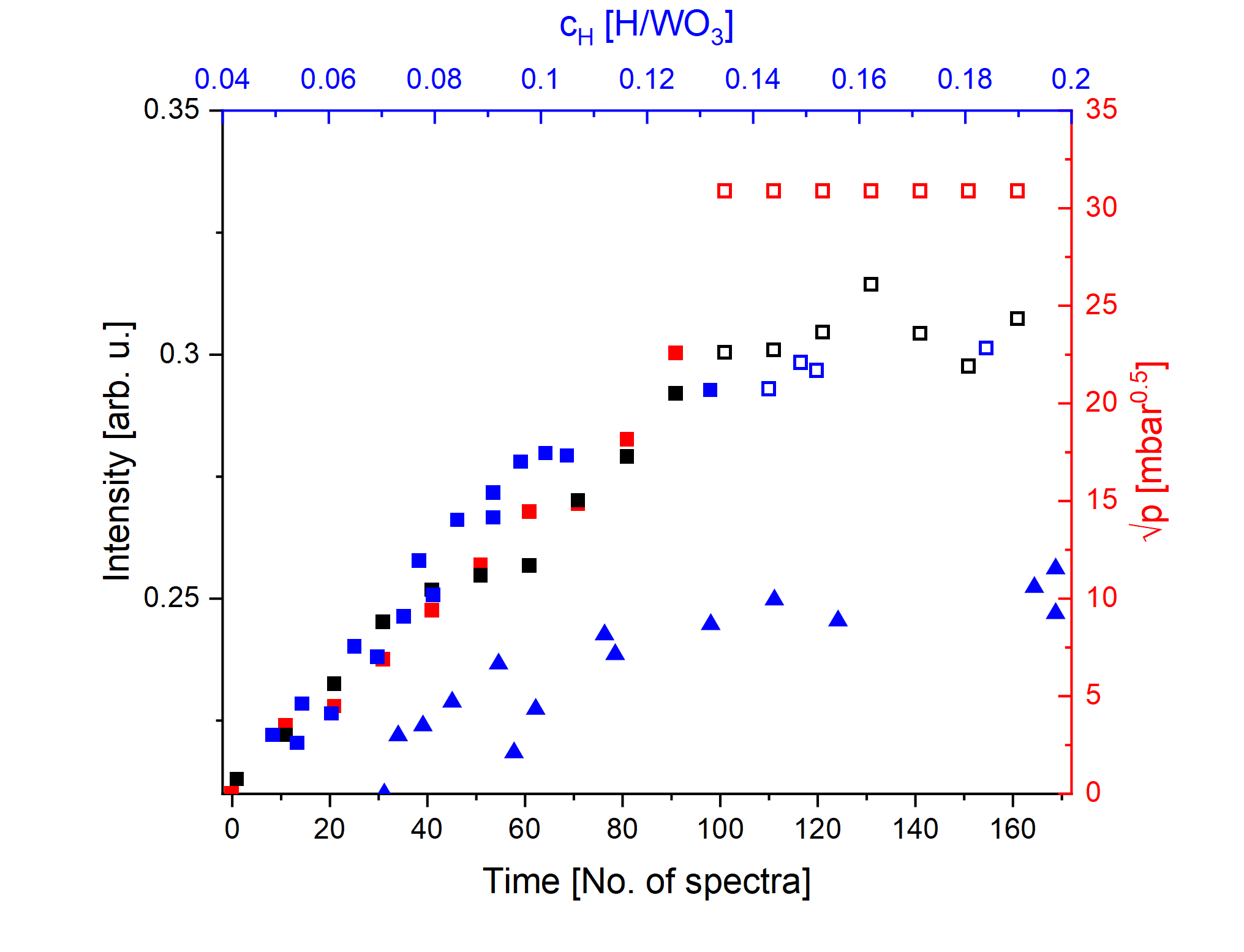}
	\caption{\label{fig:4_VB_H2_pressure} The intensity of hydrogen induced, conduction band like tungsten states (black) are shown as a function of time and with hydrogen back pressure (red). Full symbols indicate increasing hydrogen pressure, empty ones mean constant hydrogen back pressure. The bulk pressure-composition isotherm data on Pt:WO$_3$ (blue) at 423 K are plotted on the same pressure axis \cite{Berzins1983}. Squares are absorption whereas triangles are desorption data.}
\end{figure}

The measurement of valence band spectra was performed in parallel to the W4f core levels, showing subtle changes with hydrogenation as well (\ref{fig:4_VB_H2_pressure}). However, in contrast to the binding energy shifts of core levels, which can be interpreted rather easily \cite{Kerber1996, Bourque2016}, the H1s state is both core level as well as valence state, and thus not easily characterized. This challenge is tackled by an in-depth analysis of the valence band spectra with the help of DFT calculations.

\begin{figure}
	\includegraphics[width=0.5\textwidth]{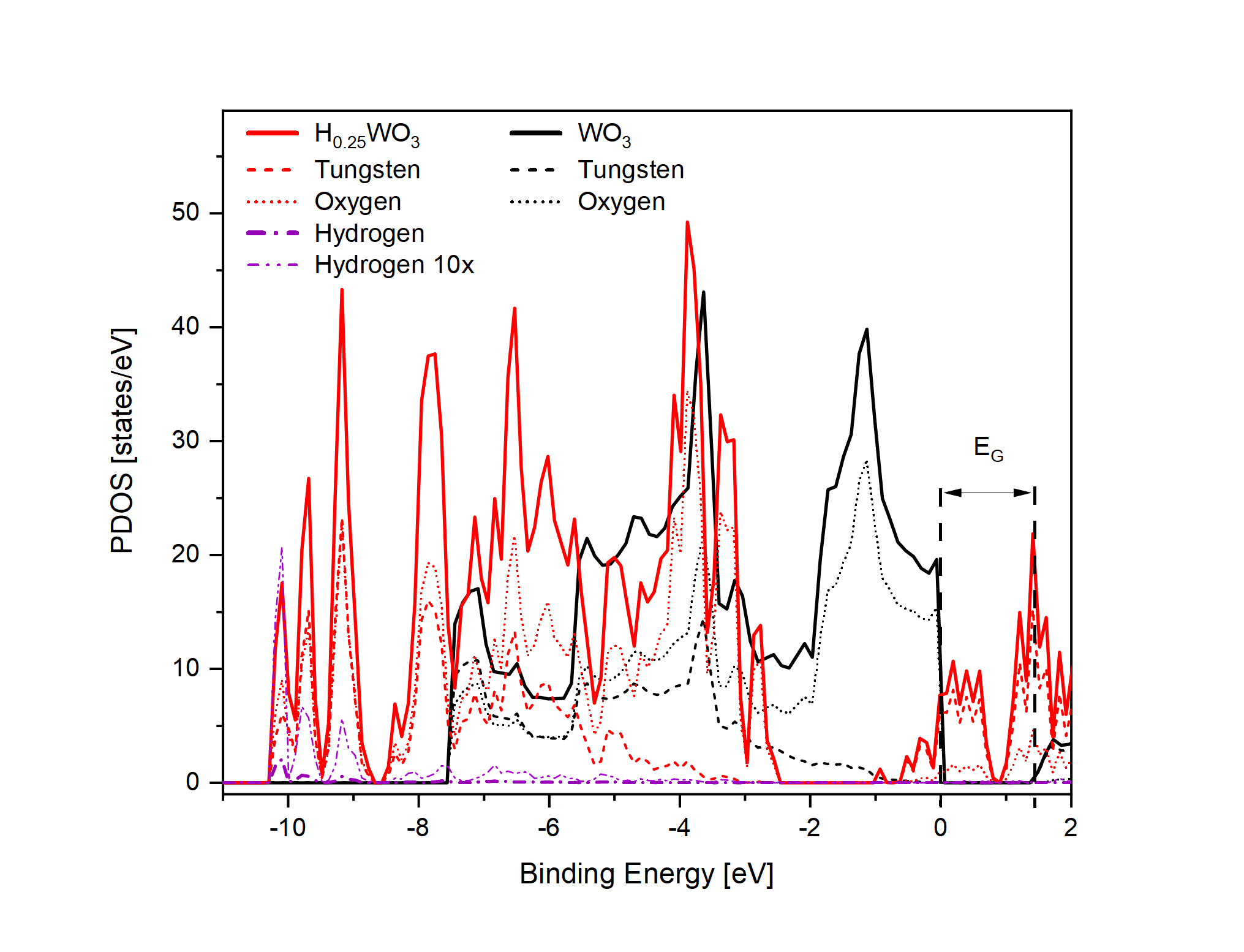}
	\caption{\label{Fig:eStructure}
	Modelled partial and total DOS of cubic WO$_3$ (black lines) and H$_x$WO$_3$ (red lines) with hydrogen concentrations as small as $x=0.25$. $E_G = 1.3$ eV is the indirect band gap in WO$_3$ derived from the onsets of valence and conduction bands.  Full band structure calculations give the direct band gap of 2.3 eV (not shown). Note the minimal contribution of hydrogen electrons (purple dashed line) to the total density of states in H$_{0.25}$WO$_3$.}
\end{figure}

DFT calculations were performed on simple cubic WO$_3$, H$_{0.25}$WO$_3$, H$_{0.75}$WO$_3$, and H$_{1}$WO$_3$ to illustrate the effects of hydrogenation on the electronic structure. WO$_3$ exhibits typical features of a semiconductor with a direct band gap of about 2.3 eV  and an indirect one of 1.3 eV as derived from band structure calculations (see Fig.~\ref{Fig:eStructure}). As discussed in section \ref{subsec:DFT}, the gap is underestimated (experimental gap of monoclinic WO$_3$ is 3.2 eV \cite{Green1991,Saenger2008}) due to the assumption of a cubic lattice, which simplifies the calculation of hydrogen intercalated WO$_3$. Nevertheless, the overall electronic structure with the valence band edge dominated by oxygen orbitals is in perfect agreement with literature. This is also in agreement with experimental observations: the optical absorption in crystalline H$_x$WO$_3$  is very similar to that in amorphous H$_x$WO$_3$ indicating that the crystallinity plays only a minor role for electronic structure changes upon hydrogenation \cite{Schirmer1980}.

\begin{figure}
	\includegraphics[width=0.5\textwidth]{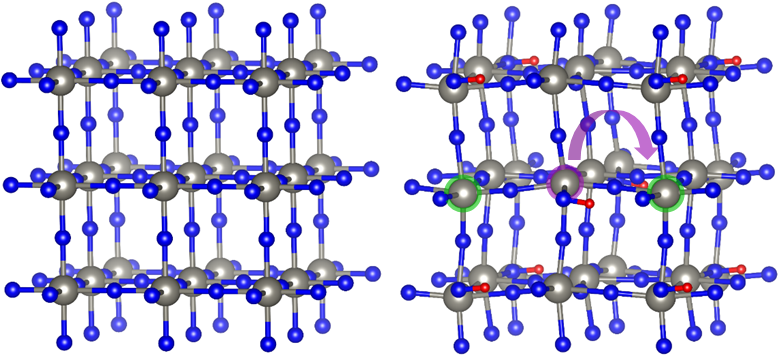}
	\caption{\label{Fig:Structure}
		Illustration of the cubic crystal structure of WO$_3$ (left) and of one polymorph of H$_{0.25}$WO$_3$ (right) as used for modelling. Note the distortion of the oxygen atoms induced by hydrogen amounts as little as $x=0.25$. This distortion is local: as an example, two tungsten atoms (grey) exclusively surrounded by oxygen atoms (blue atoms) are highlighted with green corona, one surrounded by five oxygens and one OH group (hydrogen in red) is highlighted with a purple corona. In a local picture, the latter atoms are W$^{5+}$, the ones surrounded solely by oxygen W$^{6+}$. Moving an electron and hydrogen (arrow) requires structural re-arrangements  }
\end{figure}

The intercalation of hydrogen into WO$_3$ was modelled by placing hydrogen in the center of the WO$_3$ subunit and optimizing its position in the simple cubic structure by energy minimization. The minimum was found with hydrogen occupying a position near one oxygen site with a distance of 1.04 \AA~ identical to the calculations of Hjelm et al. \cite{Hjelm1996}, and very similar to the bond length of an OH ion. Simultaneously, the lattice is slightly distorted (Fig.~\ref{Fig:Structure}). The presence of hydrogen accompanied by the lattice distortion leads to two main changes in the electronic structure (see Fig. \ref{Fig:eStructure}): the valence band broadens due to the bonding of oxygen with hydrogen (bonding states around -11 eV) and the conduction band minimum dominated by tungsten states is being occupied by electrons donated by the hydrogen and pushed below the Fermi level. It is expected that these changes lead to large increases in the electronic conductivity, which will be discussed in section~\ref{sec:discuss}. However, the aim of this paper is to confirm these effects by photoelectron spectroscopy, as theses states should be observable.

Figure~\ref{Fig:DFT_vs_Exp} shows the calculated DOS for H$_{0.25}$WO$_3$ and H$_{0.75}$WO$_3$ that have been broadened to account for the natural linewidth of the photoemission process and valence band spectra at low and high applied hydrogen back pressure. 
The effects of going from H$_{0.25}$WO$_3$ to H$_{0.75}$WO$_3$ (see Fig. \ref{Fig:DFT_vs_Exp}) are less drastic  compared to the difference between WO$_3$ and H$_x$WO$_3$ (see Fig. \ref{Fig:eStructure}). Most striking is the increasing hydrogen - oxygen bonding states and corresponding broadening of the oxygen states, which leads to a decrease of DOS near 3 eV binding energy. Second effect, most relevant to this paper is the increase of tungsten states near $E_F$ with increasing hydrogen content. These two changes  match the experimental spectra exceptionally well (Fig. \ref{Fig:DFT_vs_Exp}). In particular, the calculations confirm the existence of hydrogen induced states near $E_F$, which are associated with tungsten states.

The normalized peak area of these hydrogen induced tungsten states at 0.4~eV binding energy scales with the square root of the hydrogen back pressure (Fig. \ref{fig:4_VB_H2_pressure}). It is important to note that it does not evolve any further with time at constant pressure, which excludes its origin induced by radiation damage and confirms that the measured state is in thermodynamic equilibrium with hydrogen gas. Since the number of hydrogen induced tungsten states is directly related to the applied hydrogen back pressure by employing our membrane approach, we plot it together with the bulk pressure-composition isotherm of Pt:WO$_3$ reported in literature \cite{Berzins1983} using the common pressure axis. Using this calibration we extract that the initial hydrogen content in our thin films is  H$_{0.04}$WO$_3$ increasing to H$_{0.14}$WO$_3$ at 1000 mbar.

\section{\label{sec:discuss}Discussion}
Already in the past, Hollinger et al. proposed that coloration in amorphous tungsten oxide films upon hydrogenation is connected to an increase of localized electronic states (W$^{5+}$), but could not disentangle the origin of the effect to be from oxygen defects caused by UV-exposure or from hydrogen intercalation \cite{Hollinger1976}.   The optical absorption band corresponds to  a peak near to the Fermi level detected by photoelectron spectroscopy. 

With this, our measurements confirm most of the literature data on electronic structure changes in H$_x$WO$_3$ or substoichiometric WO$_{3-\delta}$, in particular the evolution of the peak near to the Fermi level \cite{Hollinger1976, DeAngelis1977, Bussolotti2003, Bouvard2016, Johansson2016}. However, with our experimental procedure we can disentangle the contributions from oxygen deficiency and hydrogen incorporation. The membrane photoemission measurements show that hydrogen interaction with WO$_3$ leads to band gap states without the loss of oxygen (Appendix, Fig.~\ref{fig:700eV_spectra}).

Attributing these states near to the Fermi level to hydrogen is too short-viewed. Even without sophisticated band structure calculations we can estimate the potential effect: the valence band of H$_x$WO$_3$ consists of $3\cdot 6$ O2p states (we neglect here the small contribution of W5d), and at most 0.2 hydrogens. In addition, we have to consider the photoionization cross section of electrons localized on hydrogen and oxygen of 0.02~Mb and 1~Mb, respectively, at around 104~eV \cite{Yeh1985}. The expected intensity ratio between hydrogen states and total valence states is $ 0.2 /(3\cdot 6) \cdot 0.02/1=2.2$ $\cdot 10^{-4}$. It is thus unlikely that the photoelectrons near $E_F$ stem from hydrogen. Instead, the DFT modelling shows that hydrogenation of WO$_3$ leads to the formation of OH with hydrogen states appearing at -11 eV and tungsten like conduction band states near the Fermi level (Fig.~\ref{Fig:DFT_vs_Exp}). 

Conductivity and optical experiments evidence that hydrogen intercalated WO$_3$ is not an electronic conductor at small hydrogen content in contrast to what is suggested by the calculated DOS (Fig. \ref{Fig:eStructure}), which shows states at $E_F$. This discrepancy was already found in the past, and explained by correlation models \cite{Crandall1977}. In local picture, hydrogen intercalation gives:
\begin{center}
	\ch{W^{6+} O_3^{2-} ->[ H ]  W^{5+} (OH)^- O_2^{2-}}, \hfill(1)\\
\end{center}
with the electron from the hydrogen localized by the OH bond. With this, the nearby  tungsten atom is different from the ones surrounded solely by oxygen atoms (Fig.~\ref{Fig:Structure}). This local picture is still compatible with the band structure, however, dynamic properties will depend on the correlation of the corresponding electrons. Such calculations are beyond the scope of this paper. We restrict ourselves to the following simplified picture explaining the observation that despite having electrons near the Fermi level (Fig.~\ref{Fig:DFT_vs_Exp}) H$_x$WO$_3$ does not exhibit metallic character (no Drude absorption \cite{Saenger2008}): The movement of electrons associated with the tungsten atom nearby the OH group (W$^{5+}$) is correlated with the hopping of the proton, which requires crystallographic rearrangement as the oxygen orientations around the tungsten atoms are slightly different (see Fig.~\ref{Fig:Structure}). 

Crandall and  Faughnan \cite{Crandall1977} find a metal-insulator transition around hydrogen concentrations of approximately 0.3, which is close to that predicted by percolation theory. At the percolation threshold, the hopping between the tungsten will form a continuous chain throughout the crystal.
Further evidence for the localization of these electrons was provided by optical absorption spectroscopy where amorphous H$_x$WO$_3$ exhibits an absorption maximum in the near infrared range \cite{Deb1973, Wittwer1978}. The optically induced jumping of the electron from one W site (W$_a$) to the other (W$_b$) is a special charge transfer excitation, which is well described by the so-called polaron model \cite{Saenger2008}: 
\begin{center}
	\ch{W_a^{5+} (OH) + W_b^{6+} O ->[$\hbar \omega$] W_a^{6+} O + W_b^{5+} (OH)} \hfill (2)
\end{center} 
In contrast to Saenger et al., who attribute the corresponding interaction of this electron with the surrounding distorted oxygen deficient lattice, we propose the interaction to be between the proton and the electron on W$^{5+}$ also leading to a distorted lattice. This so-called proton polaron has been proposed for describing the proton transport in the similar system, hydrated BaCe$_{0.8}$Y$_{0.2}$O$_{3- \delta}$ \cite{Braun2017}. The general underlying principle, the required crystalline rearrangements linked to charge transfer ("movement of distortions", see Fig.~\ref{Fig:Structure}, and energy dependence on hydrogen constellation, Appendix, Fig.~\ref{fig:HconstellationDFT}), remains the same. The metal oxygen rearrangement proceeds through vibrations of the WO$_6$ octahedra  and O-H vibrations. The mobility of protons may thus be in both cases a result of phonon-assisted jumps. The changes of the electronic structure by a proton-oxygen bond also explains the difference to changes induced by oxygen vacancies. Here, recent DFT calculations provide arguments that oxygen vacancies in WO$_3$ introduce 'free' carriers to the conduction band, because the corresponding electrons are delocalized over a large area \cite{Wang2016}. In our case, the OH bond pins the similar electron brought into the system by hydrogen intercalation. From an electron spectroscopy point of view, this difference is indistinguishable, but the localized picture is in good agreement with other physical properties of H$_x$WO$_{3-\delta}$ (see discussion above).

Finally, tackling the difficulties of probing the electronic structure in H$_x$WO$_{3-\delta}$ by static electron spectroscopy by following dynamic changes \textit{in-situ} may be extended to ultrafast spectroscopy. With this, the degree of localization is directly accessible.

The main outcome of this study is the change of the electronic structure near the Fermi edge from from oxygen dominated orbitals to tungsten dominated orbitals (see Fig.~\ref{Fig:eStructure}). This has drastic consequences on (electro-) catalysis of oxygen related reactions in addition to the physical properties like electron conductivity and optical properties as discussed above. WO$_3$ is one of the oxides considered to be a suitable photo electro-chemical water splitting catalysts \cite{Kalanur2018}. In particular, it shows a high activity for water oxidation in the presence of a suitable electron acceptor \cite{Yan2015,Kalanur2018}.  Hydrogen treatment of WO$_3$ can increase photocurrent significantly \cite{Yan2015,Singh2015}, which was interpreted as an increase in the number of oxygen vacancies, and thus defect states. With our findings, we cannot exclude the existence of oxygen defect states in parallel to proton polarons. However, the more pronounced effect of hydrogen treatment than vacuum annealing \cite{Yan2015,Singh2015} suggests that generally hydrogen related states are the main origin of the effect.

Similarly, the hydrogen induced electronic structure near the Fermi edge explains the rather small effect on superconductivity in tungsten oxides, if compared to metal doping \cite{Reich2009,Pellegrini2019}. If hydrogen vibrations with H-PDOS at $E_F$ contributed to the superconductivity, a strong positive effect may be expected. However, the superconductivity in bulk WO$_3$ is currently  understood to originate from  a  weak-coupling  state  sustained  by  soft  vibrational  modes  of  the  WO$_6$ octahedra \cite{Pellegrini2019}, and the formation of OH instead of the creation of oxygen vacancies upon hydrogenation is in line with this argumentation.

\section{\label{sec:4}Conclusions}
We measured the palladium assisted hydrogenation of tungsten trioxide by \textit{operando} membrane XPS at hydrogen pressures up to 1 bar. The combination of the membrane XPS with synchrotron radiation allows to measure both core levels and valence band simultaneously up to high hydrogen pressures minimizing the effect of initial preparation parameters, beam damage and exposure to the reducing UHV environment that would occur with consecutive XPS/UPS measurements. Analysis of the tungsten 4f core levels shows an increase of W$^{5+}$ with increasing hydrogen pressure. At the same time a state appears close to the Fermi level, that is connected to the hydrogen content in the compound. Although induced by hydrogen, the corresponding electrons are allocated to tungsten d-states. Using previous reference measurements it is possible to create a pressure-composition isotherm from the spectroscopy data. The photoemission measurements together with DFT calculations corroborate that the coloration of the films by hydrogen can be explained by a proton polaron model.

\begin{acknowledgments}
	This work was partly supported by the UZH-UFSP program LightChEC. Financial support from the Swiss National Science Foundation (grant number 172662 and Requip grant 182987) is greatly acknowledged. We thank Dr. C. Puglia (Uppsala University, Sweden) and the Carl Tygger Foundation for the availability of VG Scienta's SES-200 photoelectron analyzer at the GasPhase beamline.
\end{acknowledgments}

\section{Appendix}
\subsection{Experimental details of the photoemission experiments}
The table \ref{tab:XPS_Parameters} shows acquisition parameters for XPS and synchrotron photoelectron spectroscopy used.
\begin{table*}
    \centering
    \begin{tabular}{c||c|c|c|c}
         &  Energy Range & Pass Energy & Energy Stepsize & total acquisition time / point \\
        \hline \textit{ex-situ} XPS Survey & 1100-0~eV & 112~eV & 100~meV & 0.4~s \\
      \textit{ex-situ}  XPS narrow scan & variable & 55~eV & 50~meV & 1.2~s \\
         \hline \textit{in-situ} XPS Survey & 550-0~eV & 20~eV & 200~meV & 0.6~s \\
       \textit{in-situ} XPS narrow scan & variable & 20~eV & 50~meV & 0.1~s \\
        
    \end{tabular}
    \caption{Acquisition parameters for XPS and synchrotron photoelectron spectroscopy.}
    \label{tab:XPS_Parameters}
\end{table*}

\begin{figure}
	\includegraphics{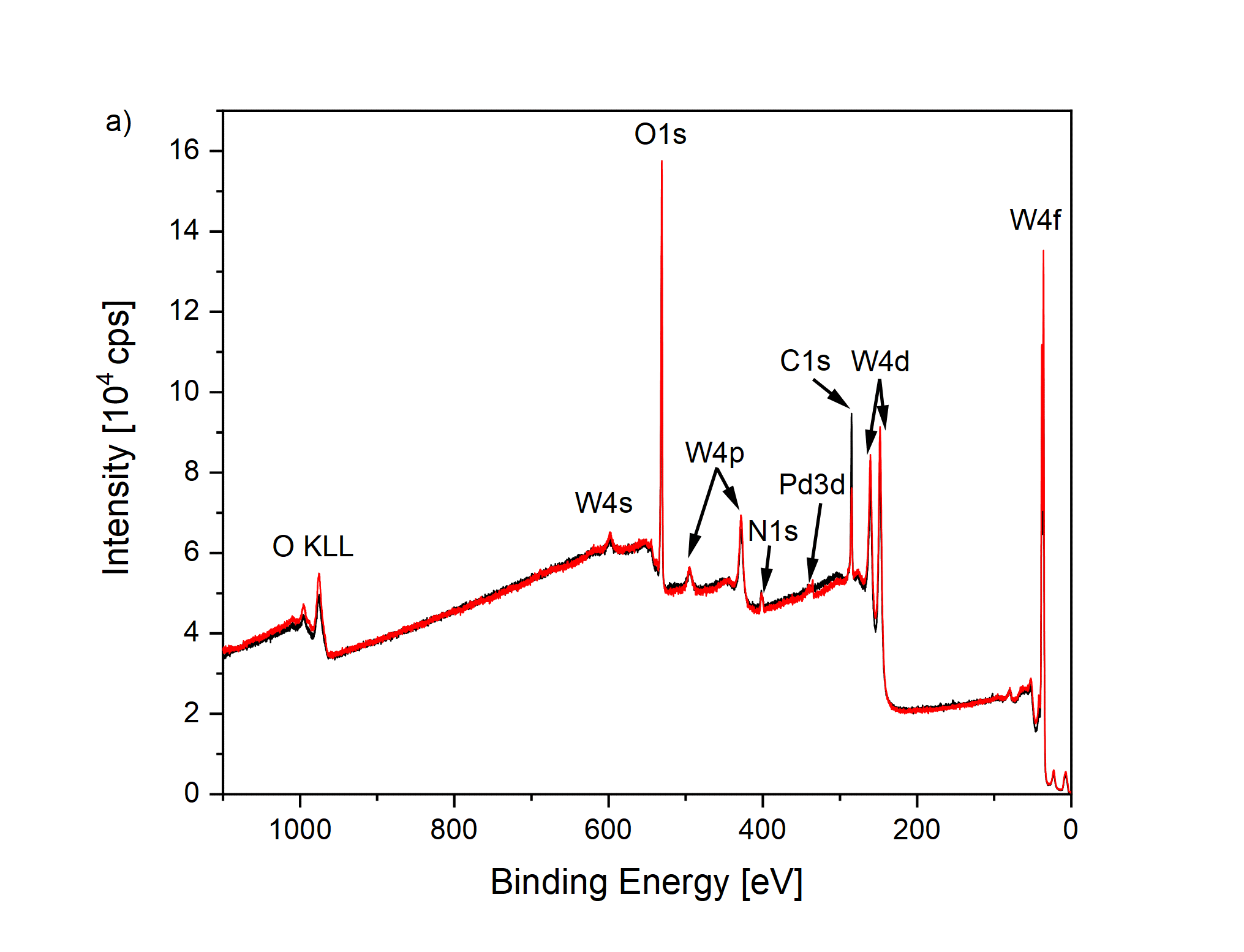}
    \includegraphics{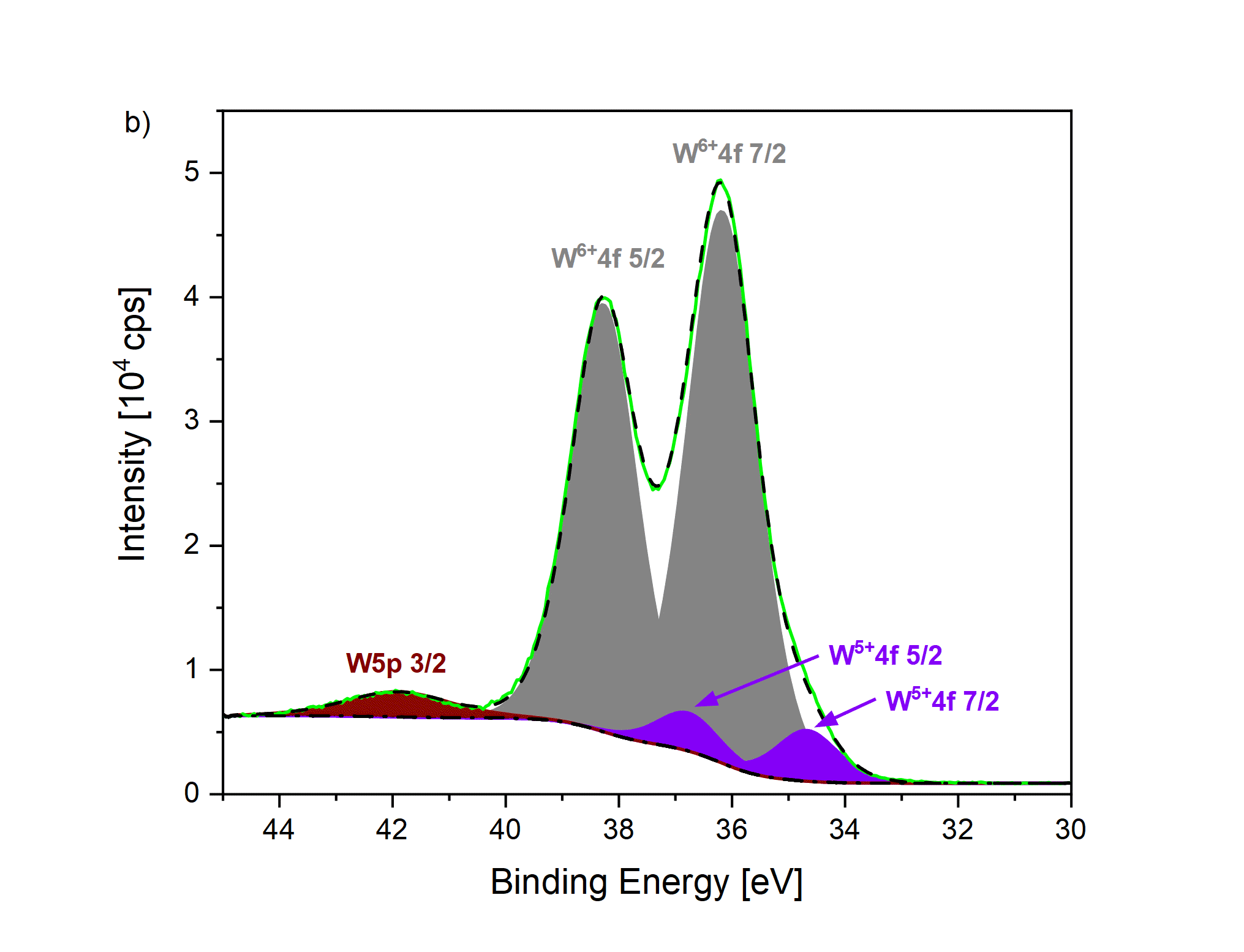}
	\caption{\label{fig:Survey_spectra} a) The pre- and post-hydrogenation XPS survey spectra are shown in black and red respectively. b)  The W5p$_{3/2}$ (brown) and W4f (grey/violet) core levels  of the electrodeposited  WO$_3$ as prepared, as described in section \ref{subsec:Sample Prep}, measured by Al K$\alpha$ XPS. The spectra are fitted with two pairs of doublets giving the W$^{6+}$ and W$^{5+}$ fractions, respectively.  The peak fitting indicates the sample being almost completely oxidized, with a small fraction of W$^{5+}$ states present ($\simeq 7.7 \%$). }
\end{figure}

\begin{figure}
    \centering
    \includegraphics{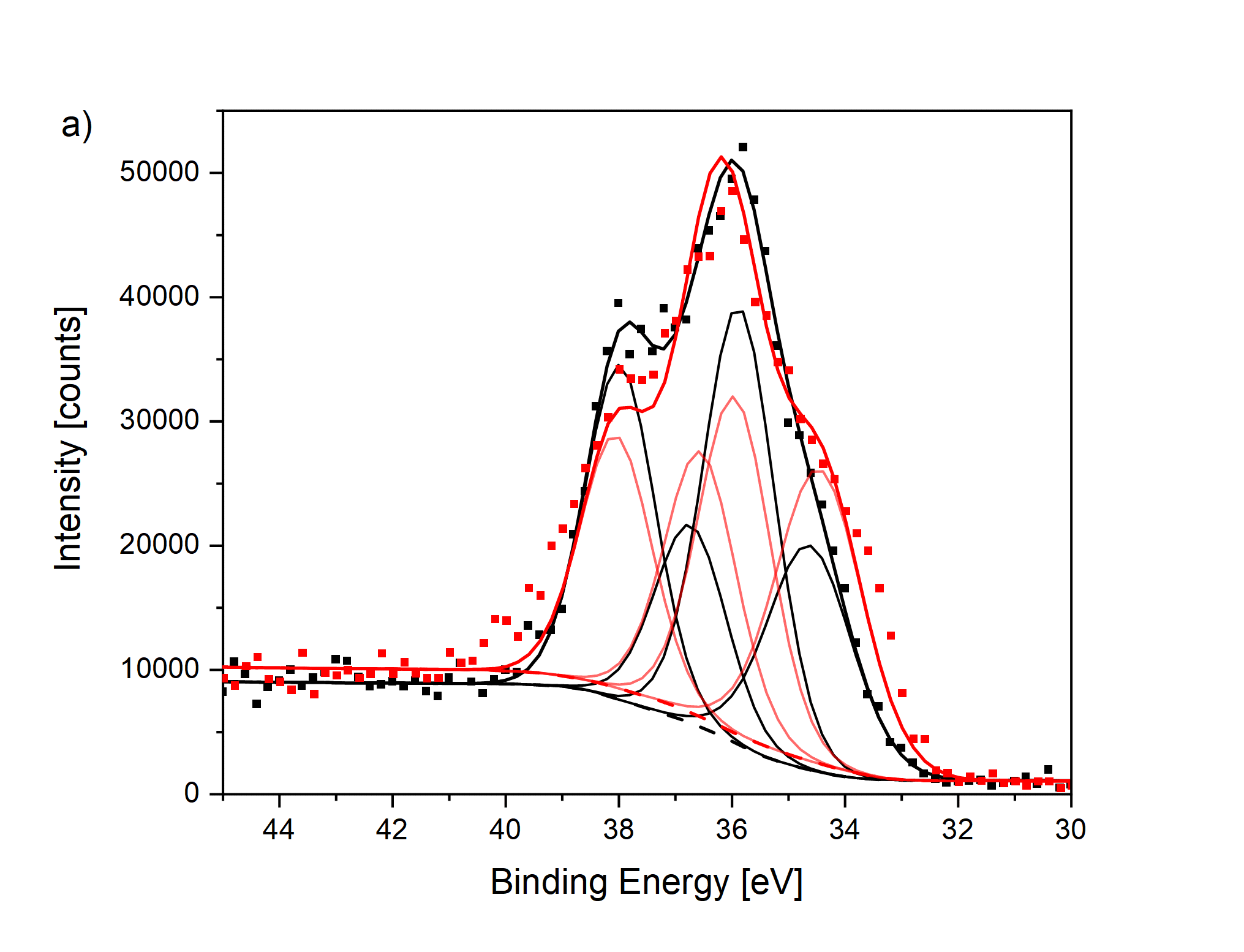}
    \includegraphics{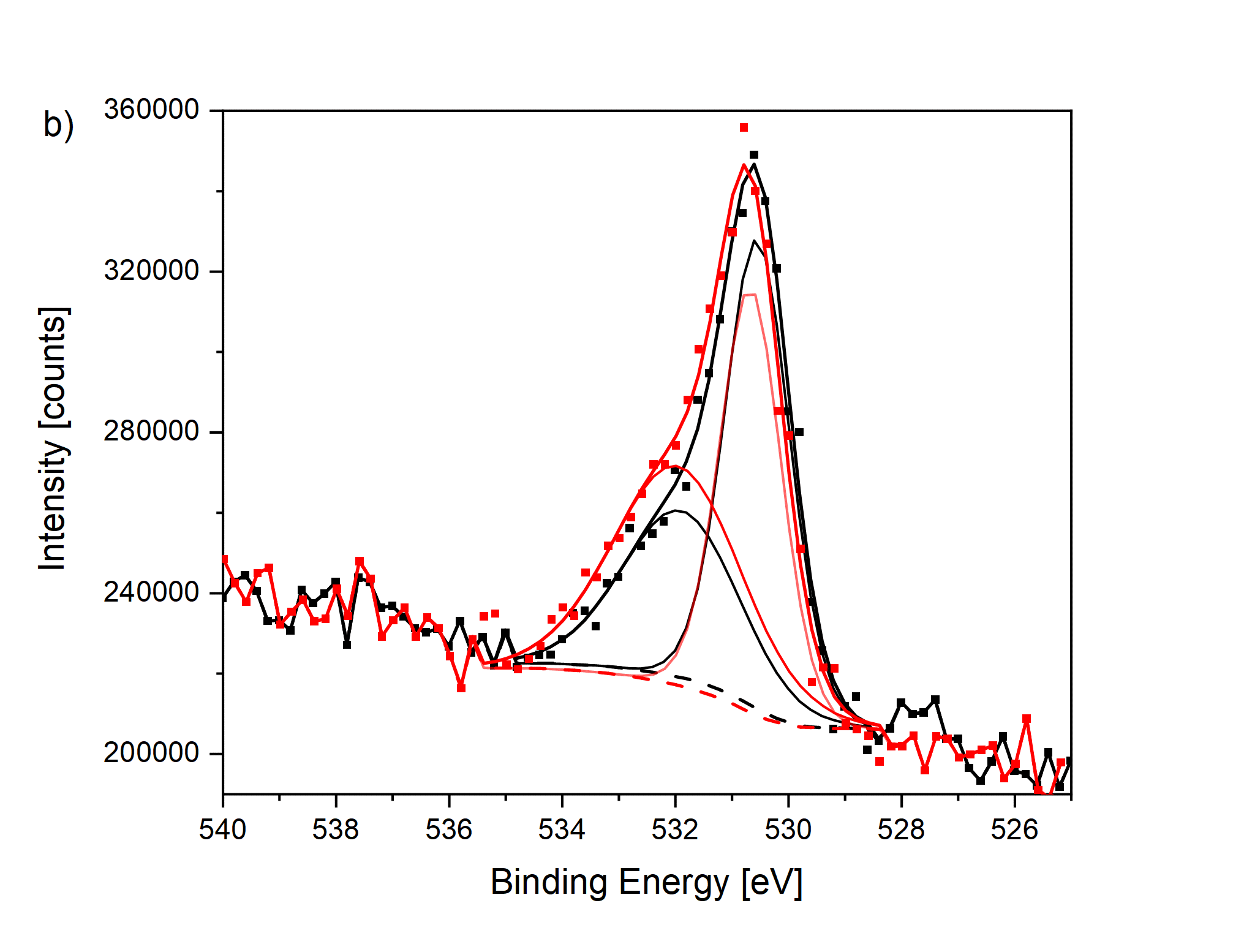}
    \caption{W4f (a) and O1s (b) core level spectra recorded using 700~eV photon energy. The spectra were recorded \textit{in-situ} before and after exposure to hydrogen (black and red respectively). The W$^{5+}$ fraction increases from 37.6\% to 51.2\% and the oxygen defect fraction from 38.8\% to 52.0\% upon hydrogenation. The intensity ratio of W to O remains nearly constant with $I(O_{latt})/I(W)= 1.06 \rightarrow 1.03 $.}
    \label{fig:700eV_spectra}
\end{figure}

Figures \ref{fig:Survey_spectra} and \ref{fig:700eV_spectra} compare the survey  and W4f core level spectra as measured \textit{ex-situ} (XPS) and \textit{in-situ} (synchrotron photoemission) before and after hydrogenation. In addition, the deconvolution of the spectra to derive the W$^{6+}$ and W$^{5+}$ fractions are shown for one archetypal example. From the intensity ratio of the O1s to W4f, one can estimate that the oxygen loss is below 3\%.

\begin{figure}
	\includegraphics{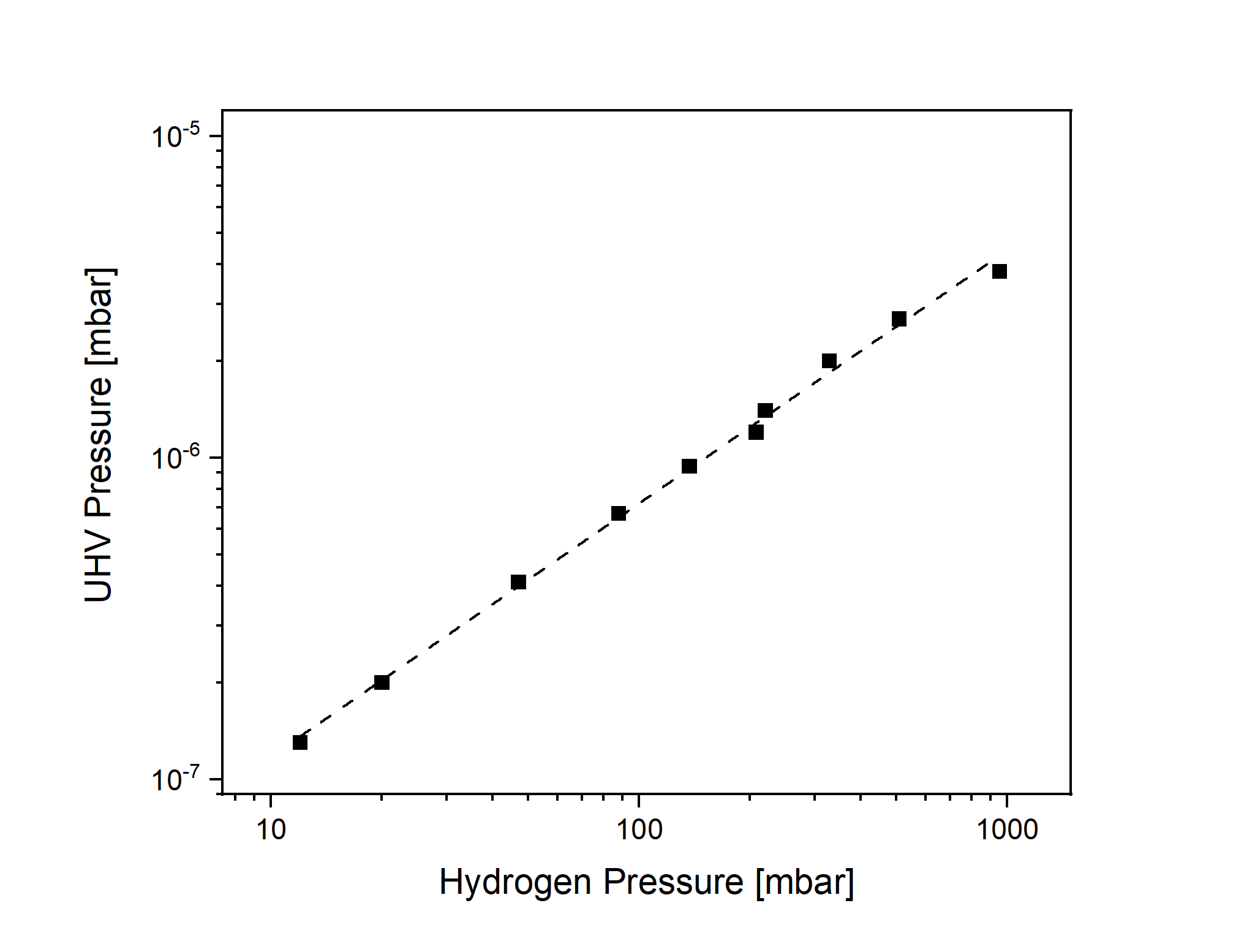}
	\caption{\label{fig:doublelog} Double-log plot of the measured hydrogen pressure in the vacuum chamber as a function of the applied feed pressure. A fit yields a slope of 0.79.  }
\end{figure}

The rate-limiting step of the permeation of hydrogen through the membrane  can be derived from the pressure dependence on both sides of the membrane: 
\begin{equation}
  p_{UHV} \propto p_{appl}^r
\end{equation}
For ideal cases, surface limited permeation gives an exponent $r=1$, bulk diffusion gives $r=\frac{1}{2}$.\cite{Delmelle2015,Sambalova2018} A fit to the double log plot of the measured hydrogen pressure in the vacuum chamber as a function of the applied feed pressure yields an exponent of 0.79 (Fig.~\ref{fig:doublelog}), indicative for surface limited kinetics.
\subsection{Hydrogen constellations used for DFT calculations}
\begin{figure}
    \centering
   \includegraphics[width=0.50\textwidth]{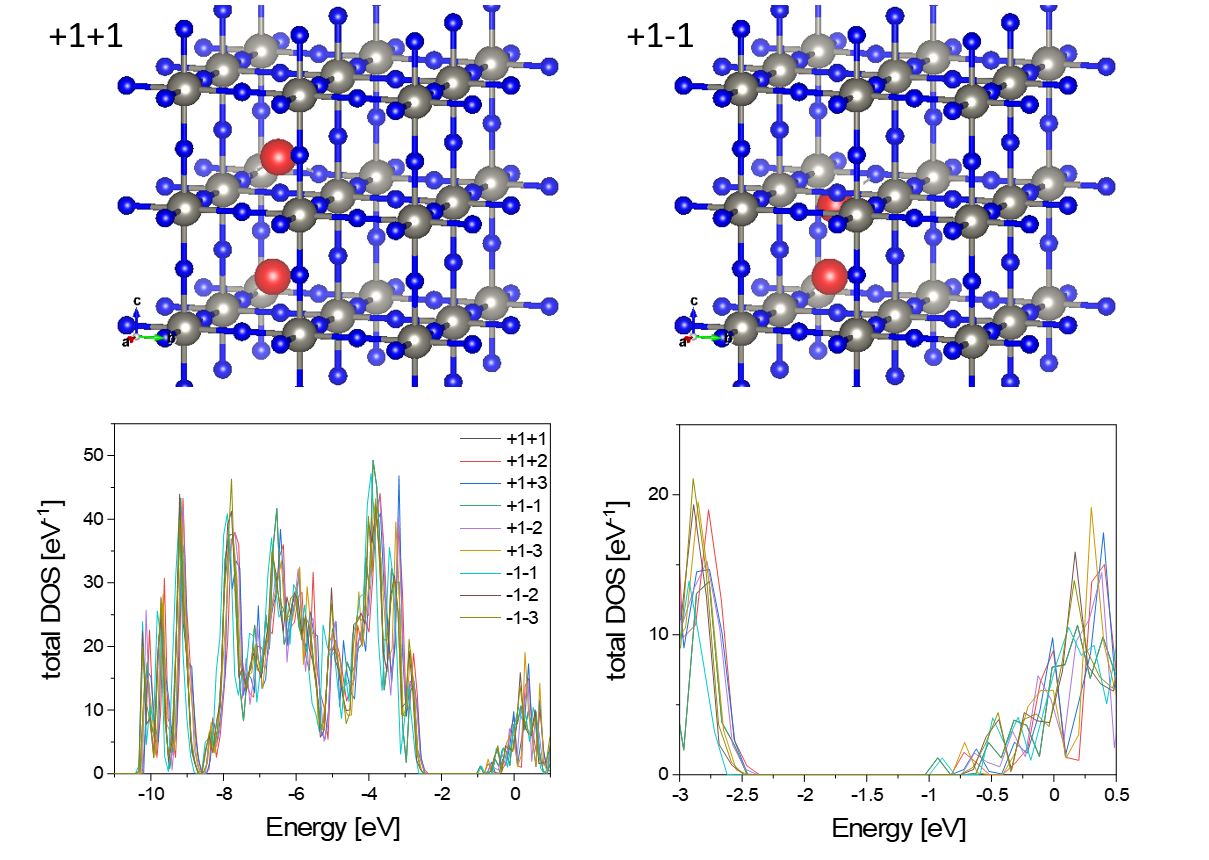}
    \caption{Top: as an example, two different hydrogen constellations in H$_{0.25}$WO$_3$, denoted as +1+1 and +1-1, before relaxation. Bottom: DFT of nine different constellations. The right graph is an enlargement around the band gap. }
    \label{fig:HconstellationDFT}
\end{figure}

Substoichiometric concentrations ($x<1$) in H$_x$WO$_3$ are difficult to model. For $x=0.25$, the unit cell of H$_{0.25}$WO$_3$ is extended to  a 2x2x2 supercell of the simple unit cell of H$_1$WO$_3$ based on one H in the WO$_3$ cell (Fig.~\ref{fig:HconstellationDFT}). Not all of the now possible H-position are occupied to reflect the substoichiometry giving additional freedom where to place these hydrogen atoms relative to each other. We modelled nine archetypal constellations, where the hydrogen atoms are placed in the subcells along the 100, 110 and 111 directions and with different spin constellations. The corresponding energy shifts are of the order of 0.3 eV. This difference is neglected within the interpretation of the photo-electron spectra, however, the non-zero energy difference will contribute to the energy barrier for hydrogen motion in H$_x$WO$_3$.

\end{document}